\documentclass[conference]{IEEEtran}
\IEEEoverridecommandlockouts
\usepackage[letterpaper, left=0.625in, right=0.625in, bottom=1.05in, top=0.75in]{geometry}
\usepackage{amssymb,amsmath}
\usepackage{graphicx}
\usepackage{cite}
\usepackage{txfonts}
\usepackage{mathrsfs} 
\usepackage{fontenc}
\usepackage[binary-units]{siunitx}
\DeclareSIUnit{\bps}{bps}
 
\usepackage[caption=false,font=footnotesize]{subfig}
\usepackage{tabularx}
\usepackage{multirow}
\usepackage{diagbox}
\usepackage{algorithm2e}
\usepackage{bbm}
\usepackage{flushend}
\usepackage{xcolor}

\begin{document}

\title{Opportunistic Reflection in Reconfigurable Intelligent Surface-Assisted Wireless Networks}

\author{\IEEEauthorblockN{Wei Jiang\IEEEauthorrefmark{1} and Hans D. Schotten\IEEEauthorrefmark{2}}
\IEEEauthorblockA{\IEEEauthorrefmark{1}German Research Center for Artificial Intelligence (DFKI)\\Trippstadter Street 122,  Kaiserslautern, 67663 Germany\\
  }
\IEEEauthorblockA{\IEEEauthorrefmark{2}Rheinland-Pf\"alzische Technische Universit\"at Kaiserslautern-Landau\\Building 11, Paul-Ehrlich Street, Kaiserslautern, 67663 Germany\\
 }
\thanks{This work was supported by the German Federal Ministry of Education and Research (BMBF) through \emph{Open6G-Hub} (Grant no.  \emph{16KISK003K}).}
}
\maketitle

\begin{abstract} 
This paper focuses on multiple-access protocol design in a wireless network assisted by multiple reconfigurable intelligent surfaces (RISs). By extending the existing approaches in single-user or single-RIS cases, we present two benchmark schemes for this  multi-user multi-RIS scenario. Inspecting their shortcomings, a simple but efficient method coined opportunistic multi-user reflection (OMUR) is proposed. The key idea is to opportunistically select the best user as the anchor for optimizing the RISs, and non-orthogonally transmitting all users' signals simultaneously. A simplified version of OMUR exploiting random phase shifts is also proposed to avoid the complexity of RIS channel estimation. 
\end{abstract}

\IEEEpeerreviewmaketitle

\section{Introduction}
Reconfigurable intelligent surface (RIS) empowers programmable propagation by manipulating a set of reflection elements \cite{Ref_jiang2023performance}. 
Since each element is passive and cheap,  RIS opens a green and cost-efficient way for sustainable performance growth. 
Most prior works focus on point-to-point communications, consisting of a base station (BS), a user, and a RIS, for the sake of analysis.  However, any practical wireless system needs to accommodate multiple users, imposing the need for designing multiple access.  
In \cite{Ref_jiang2023capacity, Ref_zheng2020intelligent_COML, Ref_jiang2023orthogonal}, the researchers investigated orthogonal and non-orthogonal multiple access (NOMA) in a single-RIS-aided system. Ding and Poor provided a simple  RIS-NOMA design in \cite{Ref_ding2020simple}. 
The authors of \cite{Ref_nadeem2021opportunistic} proposed opportunistic beamforming (OppBF) to avoid channel estimation in multi-user RIS. In \cite{Ref_jiang2023userscheduling}, user scheduling and passive beamforming for FDMA/OFDMA in a RIS system are discussed. The work in \cite{Ref_jiang2022multiuser} studied the effect of discrete phase shifts on multi-user RIS communications.  On the other hand, it is possible that a wireless system deploys multiple distributed surfaces, rather than a single RIS. For example, Zheng \textit{et al.} \cite{Ref_zheng2021doubleIRS} aimed to unveil the potential of multi-RIS-aided networks, and the effect of multiple RISs in improving spectral efficiency is studied in \cite{Ref_niu2022double}. In \cite{Ref_jiang2022intelligent}, the authors presented moving vehicular networks aided by multiple RIS surfaces installed on vehicles. Moreover, user selection in multi-RIS-aided multi-user communications is reported in \cite{Ref_jiang2023userselection}.

While the aforementioned works have deepened our understanding, how to effectively coordinate multi-user transmission in a multi-RIS-aided system is still unclear. Responding to this, this paper aims to design a multi-access scheme for its uplink transmission. The major contributions are listed as follows:
\begin{enumerate}
   \item We present two benchmark schemes called joint reflection (JR) and opportunistic reflection (OR) by extending the existing approaches in single-user or single-RIS cases.
    \item Inspecting the drawbacks of these two schemes, we propose a simple but efficient scheme coined opportunistic multi-user reflection (OMUR).  It simplifies reflection optimization of JR by selecting an opportunistic user as the anchor for the RIS optimization. Meanwhile, it is more efficient than OR by exploiting multi-user gain through non-orthogonal transmission.
    \item Inspired by \cite{Ref_nadeem2021opportunistic}, a simplified version of OMUR exploiting random phase rotations is  proposed to avoid the high overhead of RIS channel estimation.
    \item The end-to-end (E2E) channel statistics under \textit{independent but not identically distributed (i.n.i.d.)} Nakagami-\textit{m} fading is delivered, and closed-form expressions of outage probability (OP) are derived for performance analysis. 
 \end{enumerate}

\section{System Model}

\begin{figure}[!t]
    \centering
    \includegraphics[width=0.45\textwidth]{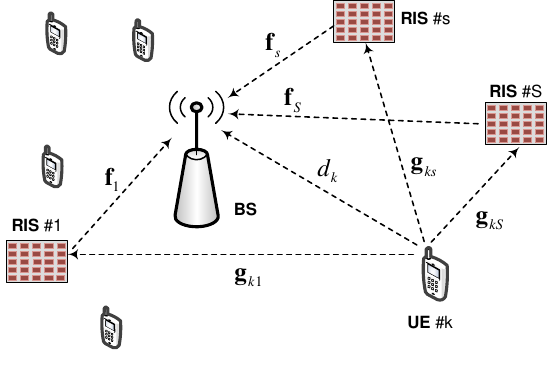}
    \caption{Schematic diagram of the uplink in a multi-RIS-aided multi-user communications system over \textit{i.n.i.d.} fading channels.  }
    \label{fig:SystemModel}
\end{figure}

This paper focuses on the uplink transmission of a wireless system, where $K\geqslant 1$ users access to a BS with the aid of $S\geqslant 1$ surfaces\footnote{The algorithm design and theoretical analysis are conducted under a generalized and more challenging setup of $S\geqslant1$. Simply setting $S=1$, it falls back to the single-RIS scenario.}, as shown in \figurename \ref{fig:SystemModel}. Each RIS is equipped with a smart controller that adaptively adjusts the phase shift of each reflecting element based on the knowledge of instantaneous channel state information (CSI).  As most of prior works such as \cite{Ref_zheng2020intelligent_COML, Ref_ding2020simple, Ref_zheng2021doubleIRS, Ref_niu2022double, Ref_wu2019intelligent},  the BS is assumed to perfectly know CSI. The $s^{th}$ surface,  $s\in \{1,2,\ldots,S\}$, is comprised of $N_s$ elements, where a typical element $n_s\in \{1,2,\ldots,N_s\}$ is modeled as $\epsilon_{sn_s}=e^{j\theta_{sn_s}}$ with a phase shift $\theta_{sn_s}\in [0,2\pi)$.  

\subsection{Independent Non-Identical Nakagami-\textit{m} Channels}
Unlike conventional wireless systems that use \textit{independent and identically distributed (i.i.d.)} Rayleigh or Rician channels, RIS-aided communications suffer from a heterogeneous fading environment. Moving users are generally surrounded by scatters, whereas the BS-RIS link is quasi-stationary with a line-of-sight (LOS) path due to fixed deployment positions. It therefore needs a more general and practical \textit{i.n.i.d.} Nakagami-$m$ fading to model RIS-aided systems.  
Write $ X \sim \mathrm{Nakagami}(m, \Omega)$ to represent a random variable (RV) following Nakagami-$m$ distribution. Its probability
density function (PDF) and cumulative distribution function (CDF) are $f_X(x)=\frac{2{m}^{m}}{\Gamma(m){\Omega}^{m}} x^{2m -1} e^{-\frac{m}{\Omega} x^2}$ and $F_X(x)=\frac{\gamma \left(m, \frac{m}{\Omega} x^2 \right)}{ \Gamma(m)}$, respectively,
where $m$ indicates the severity of small-scale fading, $\Omega$ equals large-scale fading.
Rayleigh fading is a special case of Nakagami-\textit{m} fading if $m=1$.  

Let $d_k\in \mathbb{C}$ denote the channel coefficient between user $k$ and the BS. We have $d_k=|d_k|e^{j\phi^d_{k}}$, where $\phi^d_{k}$ denotes phase and $|d_k|$ is magnitude, following Nakagami-\textit{m} fading, i.e., $|d_k| \sim \mathrm{Nakagami}\left(m^d_{k}, \Omega^d_{k} \right)$.
The channel between the $s^{th}$ RIS and the $k^{th}$ user is modeled as  $\mathbf{g}_{ks}=\Bigl[g_{k,s1},g_{k,s2},\ldots,g_{k,sN_s}\Bigr]^T$, where $g_{k,sn_s}=|g_{k,sn_s}|e^{j\phi^g_{k,sn_s}}$ is the channel coefficient between the $n_s^{th}$ element of surface $s$ and user $k$. All elements over a RIS has the same propagation path to a certain user or the BS, so they experience similar, if not identical, fading statistics. Hence, we can apply $m^g_{sk}$ and $\Omega^g_{sk}$, $\forall s,k$ for the channels from RIS to users, i.e. $|g_{k,sn_s}| \sim \mathrm{Nakagami}\left(m^g_{sk}, \Omega^g_{sk}\right)$, $\forall n_s$.
Similarly, write $\textbf{f}_s =[f_{s1},f_{s2},\ldots,f_{sN_s}]^T$ to denote the channel vector from the $s^{th}$ surface to the BS, where $|f_{sn_s}| \sim \mathrm{Nakagami}\left(m^f_s, \Omega^f_s\right)$, $\forall n_s$. 
\subsection{Signal Models}
The users simultaneously send their signals towards the BS over the same time-frequency resource. Let $s_k\in \mathbb{C}$ denote the information symbol from user $k$, satisfying $\mathbb{E}[|s_k|^2]\leqslant P_k$, where $P_k$ denotes the power constraint of user $k$ and $\mathbb{E}[\cdot]$ is expectation. The BS observes the received signal of
\begin{equation} \label{eqn_systemModel}
    r=\sum_{k=1}^{K} \Biggl(\sum_{s=1}^{S} \sum_{n_s=1}^{N_s}  f_{sn_s} e^{j\theta_{sn_s}} g_{k,sn_s}  + d_{k} \Biggr) s_k + n,
\end{equation}
where $n$ is additive white Gaussian noise (AWGN) with zero mean and variance $\sigma_n^2$, i.e. $n\sim \mathcal{CN}(0,\sigma_n^2)$. Alternatively, \eqref{eqn_systemModel} can be rewritten in matrix form as
\begin{equation} \label{EQN_IRS_RxSignal_Matrix}
    r=\sum_{k=1}^{K} \Biggl(\mathbf{f}^T \boldsymbol{\Theta} \mathbf{g}_{k}  + d_{k} \Biggr) s_k + n,
\end{equation}
by defining $\mathbf{g}_k=\left[\mathbf{g}_{k1}^T,\mathbf{g}_{k2}^T,\ldots,\mathbf{g}_{kS}^T \right]^T$, $\mathbf{f}=\left[\mathbf{f}_{1}^T,\mathbf{f}_{2}^T,\ldots,\mathbf{f}_{S}^T \right]^T$, and an overall phase-shift matrix $\boldsymbol{\Theta}=\mathrm{diag}\biggl\{\boldsymbol{\Theta}_1,\ldots,\boldsymbol{\Theta}_S\biggr\}$, where $\boldsymbol{\Theta}_s=\mathrm{diag}\{e^{j\theta_{s1}},\ldots,e^{j\theta_{sN_s}}\}$, $\forall s$.

\section{Multiple Access for Multi-RIS-Aided Systems}

From the information-theoretic perspective, orthogonal multiple access is inefficient because each user utilizes only a fraction of the available time-frequency resource. Hence, non-orthogonal transmission and successive interference cancellation (SIC) are applied. This yields the \textit{sum capacity} of 
\begin{equation} \label{EQN_SumRate2}
    C =\log \left(  1 +\frac{ \sum_{k=1}^K \left|\mathbf{f}^T \boldsymbol{\Theta} \mathbf{g}_k +d_{k}\right|^2P_k}{\sigma_n^2}  \right).
\end{equation}
Let $\mathbf{G}=\left[\mathbf{g}_{1},\mathbf{g}_{2}, \ldots, \mathbf{g}_{K}\right]$ and $\mathbf{d}=[d_1,d_2,\ldots,d_K]^T$. It is not difficult to verify that $\sum_{k=1}^K \left|\mathbf{f}^T \boldsymbol{\Theta} \mathbf{g}_k +d_{k}\right|^2=\left\|\mathbf{f}^T \boldsymbol{\Theta} \mathbf{G} +\mathbf{d}^T\right\|^2$.  Assume all users have the same power constraint, i.e., $P_k=P_u$, $\forall k$, \eqref{EQN_SumRate2} is equivalent to
\begin{equation} \label{eqn_IRS_sumRate}
    C=   \log  \left(  1+\frac{  \left\|\mathbf{f}^T \boldsymbol{\Theta} \mathbf{G} +\mathbf{d}^T \right\|^2P_u}{\sigma_n^2}  \right),
\end{equation}
which is a function of $\boldsymbol{\Theta}$, resulting in the optimization formula
\begin{equation}  
\begin{aligned} \label{EQN:sumRate_Optimization}
\max_{\boldsymbol{\Theta}} \quad & \left\|\mathbf{f}^T \boldsymbol{\Theta} \mathbf{G} +\mathbf{d}^T\right\|^2\\
\textrm{s.t.} \quad & \theta_{sn_s}\in [0,2\pi), \: \forall s, n_s.
\end{aligned}
\end{equation}

\subsection{Ideal Reflection}
Suppose there exist 'magic' RIS surfaces for ideal reflection  (IR), enabling optimal phase alignment $\theta_{sn_s}=\phi^d_k-(\phi^g_{k,sn_s}+\phi^f_{sn_s})$ for all users simultaneously, the maximization of \eqref{EQN_SumRate2} is achieved as $C_{ir}=\log\left(1+\gamma_{ir} \frac{P_u}{\sigma_n^2}\right)$
with the E2E channel power gain of
\begin{equation} \label{IRS_EQN_SNR_ideal}
    \gamma_{ir}=\sum_{k=1}^K \left|\sum_{s=1}^{S} \sum_{n_s=1}^{N_s}  |f_{sn_s}| |g_{k,sn_s}|  +|d_k| \right|^2.
\end{equation}
Unfortunately, it is impossible that a single value of $\theta_{sn_s}$ is optimal for different users simultaneously because $\phi^d_k\neq \phi^d_{k'}$ and $\phi^g_{k,sn_s}\neq \phi^g_{k',sn_s}$, when $k\neq k'$. Although IR does not exist, it  provides a theoretical value to indicate the performance bound. 
\subsection{Joint Reflection}
In contrast to the \textit{optimal} phase shifts of IR, a set of \textit{optimized} phase shifts can be obtained by solving \eqref{EQN:sumRate_Optimization}. To this end, we apply the semidefinite relaxation (SDR) approach, which has been used to solve a non-convex quadratically constrained quadratic program (QCQP) in a single-user single-RIS system \cite{Ref_wu2019intelligent}. 
Define $\mathbf{q}=\left[q_{1},q_{2},\ldots,q_{M}\right]^H$, where $M$ is the total number of elements over all surfaces, i.e., $M=\sum_{s=1}^SN_s$, and $q_{m}=e^{j\theta_{sn_s}}$ with $m=\sum_{s'=1}^{s-1}N_{s'}+n_s$. Let  $\boldsymbol{\chi}=\mathrm{diag}(\mathbf{f})\mathbf{G}\in \mathbb{C}^{M\times K}$, we have $\mathbf{f}^T \boldsymbol{\Theta} \mathbf{G}=\mathbf{q}^H\boldsymbol{\chi}\in \mathbb{C}^{1\times K} $.
Thus, $\left\|\mathbf{f}^T \boldsymbol{\Theta} \mathbf{G} +\mathbf{d}^T\right\|^2=\left\|\mathbf{q}^H\boldsymbol{\chi} +\mathbf{d}^T\right\|^2$, \eqref{EQN:sumRate_Optimization} is rewritten as 
\begin{equation}  
\begin{aligned} \label{EQN_IRS_QCQPoptimization}
\max_{\mathbf{q}} \quad & \mathbf{q}^H\boldsymbol{\chi}\boldsymbol{\chi}^H\mathbf{q}+\mathbf{q}^H\boldsymbol{\chi}\mathbf{d}^*+\mathbf{d}^T\boldsymbol{\chi}^H\mathbf{q}+\|\mathbf{d}\|^2\\
\textrm{s.t.} \quad & |q_m|^2=1, \:  \forall m=1,\ldots,M,
\end{aligned}
\end{equation} 
which is a QCQP. Introducing an auxiliary variable $t$, \eqref{EQN_IRS_QCQPoptimization} can be homogenized as 
\begin{equation} \begin{aligned} \label{eqn_IRS_relaxedOptimization}
    \max_{\mathbf{q}}  \quad & \left\|t\mathbf{q}^H\boldsymbol{\chi} +\mathbf{d}^T\right\|^2\\
     =\max_{\mathbf{q}} \quad & t^2\mathbf{q}^H\boldsymbol{\chi}\boldsymbol{\chi}^H\mathbf{q}+t\mathbf{q}^H\boldsymbol{\chi}\mathbf{d}^*+t\mathbf{d}^T\boldsymbol{\chi}^H\mathbf{q}+\|\mathbf{d}\|^2.
     \end{aligned}
\end{equation}
Define $\mathbf{C}=\begin{bmatrix}\boldsymbol{\chi}\boldsymbol{\chi}^H&\boldsymbol{\chi}\mathbf{d}^H\\ \mathbf{d}\boldsymbol{\chi}^H& \|\mathbf{d}\|^2\end{bmatrix},\:\:\mathbf{v}= \begin{bmatrix}\mathbf{q}\\ t\end{bmatrix}$, and $\mathbf{V}=\mathbf{v}\mathbf{v}^H$, we have $\mathbf{v}^H\mathbf{C}\mathbf{v}=\mathrm{Tr}(\mathbf{C}\mathbf{V})$, where $\mathrm{Tr}(\cdot)$ denotes the trace of a matrix.
Thus, \eqref{eqn_IRS_relaxedOptimization} is reformulated as 
\begin{equation}  \label{RIS_EQN_optimizationTrace}
\begin{aligned} \max_{\mathbf{V}}\quad &  \mathrm{Tr} \left(  \mathbf{C}\mathbf{V} \right)\\
\textrm{s.t.}  \quad & \mathbf{V}_{m,m}=1, \: \forall m=1,\ldots,M \\
  \quad & \mathbf{V}\succ 1
\end{aligned},
\end{equation}
where $\mathbf{V}_{m,m}$ means the $m^{th}$ diagonal element of $\mathbf{V}$, and $\succ$ stands for a positive semi-definite matrix.
The optimization finally becomes a semi-definite program, whose globally optimal solution $\mathbf{V}^\star$ can be got by a numerical algorithm named CVX \cite{cvx}.

\subsection{Opportunistic Reflection}
In wireless systems, opportunistic communications achieve a well performance-complexity trade-off by exploiting multi-user diversity gain. Inspired by this, we can select a user as the unique transmitter such that the reflection optimization is simplified. 
The \textit{single-user bound} for a typical user $k$, which means the capacity of a point-to-point link with the other users absent from the system,  is given by 
\begin{equation} \label{EQN_IRS_SingleUserBound}
  C_k = \log  \left(  1+\frac{\left|\mathbf{f}^T \boldsymbol{\Theta} \mathbf{g}_k +d_{k}\right|^2P_u}{\sigma_n^2}  \right),\: \forall\: k.
\end{equation}
The optimal reflection to maximize $C_k$ is $\mathbf{\Theta}_{k}=\mathrm{diag}\left \{e^{j\left( \arg\left( d_k\right)  -\arg\left( \mathrm{diag}(\mathbf{f})\mathbf{g}_k\right)\right)}\right\}$, achieving coherent combining at the receiver. The resultant capacity equals to $C_k =\log\left(1+\gamma_k\frac{P_k}{\sigma_n^2} \right)$ with
the maximized E2E channel power gain for user $k$
\begin{equation} \label{EQN_RIS_GammaK}
   \gamma_k =   \left| \sum_{s=1}^{S} \sum_{n_s=1}^{N_s}  |f_{sn_s} || g_{k,sn_s}|  + |d_{k}| \right|^2.
\end{equation}
The philosophy of OR is to determine the best user as $k^\star=\arg \max_{k\in\{1,\ldots,K\}} \left(\gamma_k\right)$.
Other users turn off when $k^\star$ is transmitting its signal, while the RIS phase shifts are tuned to $\mathbf{\Theta}_{k^\star}$.
The effective E2E power gain of OR is equal to \begin{equation}  \label{eqn_E2EchannelGain}
  \gamma_{or} = \max_{k\in\{1,\ldots,K\}} \biggl(\gamma_k\biggr)=\max_{k\in\{1,\ldots,K\}} \left (  \left| \sum_{s=1}^{S} \sum_{n_s=1}^{N_s}  |f_{sn_s} || g_{k,sn_s}|  + |d_{k}| \right|^2 \right).
\end{equation}

\subsection{Opportunistic Multi-User Reflection}

\SetKwComment{Comment}{/* }{ */}
\RestyleAlgo{ruled}
\begin{algorithm} 
\caption{Opportunistic Multi-User Reflection} \label{alg:IRS002}
\ForEach{Coherence Interval}{
Estimate $\mathbf{f}$, $\mathbf{d}$, and $\mathbf{g}_k$, $\forall k$\;
$k^\star \gets \arg \max_{k\in\{1,\ldots,K\}} \bigl( \gamma_k \bigr)$\;
Tune RISs to $\mathbf{\Theta}_{omur}=e^{j\left( \arg\left( d_k^\star\right)  -\arg\left( \mathrm{diag}(\mathbf{f})\mathbf{g}_k^\star\right)\right)}$\;
All users simultaneously transmit\;
The receiver conducts SIC\;
}
\end{algorithm}

\SetKwComment{Comment}{/* }{ */}
\RestyleAlgo{ruled}
\begin{algorithm} 
\caption{OMUR with Random Phase Shifts} \label{alg:OMUR-RP}
\ForEach{Transmission Block}{
RISs randomly set phase shifts\;
All users simultaneously transmit\;
The receiver conducts SIC\;}
\end{algorithm}

JR suffers from high computational complexity, blocking its use in practical systems. OR simplifies the optimization process, but it does not fully utilize the resource. Responding to this, we propose a simple but more efficient scheme coined opportunistic multi-user reflection. The main idea is to opportunistically select a user as the anchor for optimizing the RIS reflection. The best user is determined as $k^\star=\arg \max_{k\in\{1,\ldots,K\}} \left(\gamma_k\right)$.  The RIS phase shifts are adjusted to $\mathbf{\Theta}_{omur}= e^{j\left( \arg\left( d_{k^\star}\right)  -\arg\left( \mathrm{diag}(\mathbf{f})\mathbf{g}_{k^\star}\right)\right)}$. Unlike OR, which merely assigns the resource to a unique user, all OMUR users simultaneously transmit their signals towards the BS for a full resource usage. In this case, the BS observes 
\begin{align} 
    r_{omur}&=\sum_{k=1}^{K} \Biggl(\mathbf{f}^T \boldsymbol{\Theta}_{omur} \mathbf{g}_{k}  + d_{k} \Biggr) s_k + n\\ \nonumber
    &=\underbrace{\Biggl(\mathbf{f}^T \boldsymbol{\Theta}_{omur} \mathbf{g}_{k^\star}  + d_{k^\star} \Biggr) s_{k^\star}}_{\text{Optimal\:Reflection}}+\underbrace{\sum_{k\neq k^\star} \Biggl(\mathbf{f}^T \boldsymbol{\Theta}_{omur} \mathbf{g}_{k}  + d_{k} \Biggr) s_k}_{\text{Non-aligned\:Reflection}} + n.
\end{align}
The receiver first detects $s_{k^\star}$ by treating non-aligned reflecting signals as noise.  
Applying SIC \cite{Ref_jiang20226G}, the receiver cancels the corresponding interference $(\mathbf{f}^T \boldsymbol{\Theta}_{omur} \mathbf{g}_{k^\star}  + d_{k^\star}) s_{k^\star}$ from $r_{omur}$ and then detects the symbol with the second largest SNR. This process iterates until all symbols are detected.  It is not hard to derive that the achievable sum rate equals $C_{omur} =
    \log \left(  1 +\gamma_{omur} \frac{  P_u}{\sigma_n^2}  \right)$
with 
\begin{equation} \label{eqn_OMUR_gamma}
    \gamma_{omur}=\left| \sum_{s=1}^{S} \sum_{n_s=1}^{N_s}  |f_{sn_s} || g_{k^\star,sn_s}|  + |d_{k^\star}| \right|^2+\sum_{k\neq k^\star} \left|\mathbf{f}^T \boldsymbol{\Theta}_{omur} \mathbf{g}_k +d_{k}\right|^2.
\end{equation}
The proposed scheme is also depicted in Algorithm 1. 

Under fast fading, a quick change of CSI imposes a high overhead on CSI acquisition. Inspired by opportunistic beamforming \cite{Ref_nadeem2021opportunistic}, we further propose a simplified version of OMUR, named OMUR with random phase (OMUR-RP) to avoid the high complexity of RIS channel estimation. Without CSI, the RIS elements randomly set their phase shifts.  All users simultaneously transmit their signals and the receiver applies SIC to deal with inter-user interference, as described by Algorithm \ref{alg:OMUR-RP}.

\section{Performance Analysis}
This section analyzes and compares the performance of different schemes with respect to outage probability, defined as $P(R_0)= \mathbb{P} \left\{ \log_2(1+ \gamma ) < R_0 \right\}$, where $\mathbb{P}$ denotes mathematical probability and $R_0$ is the target rate. The magnitude of the E2E channel gain for user $k$ represented by $A_k=\sqrt{\gamma_k}=\sum_{s=1}^S\sum_{n_s=1}^{N_s} |f_{sn_s}||g_{k,sn_s}|+|d_k|$ is first statistically modeled for the sake of getting closed-form OP expressions. 

\newtheorem{lemma}{Lemma}
\begin{lemma}
\label{Lemma_01}
The distribution of $A_k$ can be approximated by \textit{the Gamma distribution}, which is characterized by the CDF 
\begin{equation} \label{eqn_e2e_mag_CDF}
	F_{A_k} (x) \approx \frac{ \gamma (\alpha_k, \beta_k x) }{ \Gamma(\alpha_k) },\: x \geq 0,
\end{equation}
with a shape parameter ${\alpha_{k}} = \frac{\left(\mathbb{E}[A_k]\right)^2}{\mathrm{Var}[A_k]} = \frac{[\mu_k(1)]^2}{\mu_k(2) - [\mu_k(1)]^2}$ and a rate parameter $	{\beta_{k}} = \frac{\mathbb{E}[A_k]}{\mathrm{Var}[A_k]} = \frac{\mu_k(1)}{\mu_k(2) - [\mu_k(1)]^2}$, where $\mathrm{Var}[\cdot]$ denotes variance.
 The first and second moment of $A_k$ are given by
\begin{align} \nonumber
	\mu_k (1) &= \frac{\Gamma(m_k^d + 1/2)}{\Gamma(m_k^d)} \left(\frac{m_k^d}{\Omega_k^d}\right)^{-1/2} + \sum_{s=1}^{S} \sum_{n_s=1}^{N_s} U_{sn_s}(1)  \\ \nonumber
	\mu_k (2) &= \Omega_k^d  + 2 \frac{\Gamma(m_k^d + 1/2)}{\Gamma(m_k^d)} \left(\frac{m_k^d}{\Omega_k^d}\right)^{-1/2}  \sum_{s=1}^{S} \sum_{n_s=1}^{N_s} U_{sn_s}(1) \\
	&+ \sum_{s=1}^{S} \left\{\sum_{n_s=1}^{N_s} U_{sn_s} (2) + 2 \sum_{n_s=1}^{N_s} U_{sn_s} (1) \sum_{n_s'=n_s+1}^{N_s} U_{sn_s'}(1) \right\} \nonumber \\
    &+ 2\sum_{s=1}^S \left\{\sum_{n_s=1}^{N_s}U_{sn_s}(1) \sum_{s'=s+1}^{S}  \sum_{n_s=1}^{N_{s'}} U_{s'n_s}(1)\right\}
\end{align}
with an intermediate function for $a=1, 2$ 
\begin{align} 
	U_{sn_s} (a)	=
	\left( \sqrt{ \frac{m^g_{sk}}{\Omega^g_{sk}} \frac{m^f_{s}}{\Omega^f_{s}} } \right)^{-a} \frac{ \Gamma(m^g_{sk} + a/2) \Gamma(m^f_{s} + a/2) }{ \Gamma(m^g_{sk}) \Gamma(m^f_{s}) }.
\end{align}
\end{lemma}
\begin{IEEEproof}
We employ a similar derivation approach as used for Theorem 1 of \cite{Ref_do2021multiRIS} to obtain the CDF of per-user distribution. The details are neglected due to the page limit.
\end{IEEEproof}

For a special case when $K=1$, i.e., a multi-RIS-aided single-user (SU) system, its E2E magnitude is defined as $A_1=\sum_{s=1}^S\sum_{n_s=1}^{N_s} |f_{sn_s}||g_{1,sn_s}|+|d_1| $.  Its OP is computed by 
\begin{align} \nonumber 
    P_{su}(R_0)&= \mathbb{P} \left\{ \log_2(1+ A_1^2P_u/\sigma^2_n ) < R_0 \right\} \\
    &=\mathbb{P}\left(A_1^2<\gamma_0/\bar{\gamma_t}\right)=\mathbb{P}\left(A_1<\sqrt{\gamma_0/\bar{\gamma_t}}\right),
\end{align} given the notations $\gamma_0=2^{R_0}-1$ and $\bar{\gamma_t}=P_u/\sigma^2_n$. Applying $x=\sqrt{\gamma_0/\bar{\gamma_t}}$ in \eqref{eqn_e2e_mag_CDF} yields $F_{A_1}\left (\sqrt{\gamma_0/\bar{\gamma_t}}\right)$, which is equivalent to $\mathbb{P}\left(A_1<\sqrt{\gamma_0/\bar{\gamma_t}}\right)$. Thus, we get
\begin{equation} \label{Eqn:SU:OutageProb}
    P_{su}(R_0)=F_{A_1}\left (\sqrt{\gamma_0/\bar{\gamma_t}}\right).
\end{equation}

For OR, its outage probability is equal to $P_{or}(R_0)=\mathbb{P} \left\{ \log_2(1+ \gamma_{or}\bar{\gamma_t} ) < R_0 \right\} = \mathbb{P}(\gamma_{or}<\gamma_0/\bar{\gamma_t})$. This yields, by recalling \eqref{eqn_E2EchannelGain},
\begin{align} \nonumber \label{Eqn:OP_OpportuntisticReflection}
    P_{or}(R_0)&= \mathbb{P}\left(\max_{k}(\gamma_{k})<\gamma_0/\bar{\gamma_t}\right)= \prod_{k=1}^{K} \mathbb{P}\left(\gamma_k<\gamma_0/\bar{\gamma_t}\right) \\ 
    &= \prod_{k=1}^{K} \mathbb{P}\left(A_k<\sqrt{\gamma_0/\bar{\gamma_t}}\right) = \prod_{k=1}^{K} F_{A_k}\left(\sqrt{\gamma_0/\bar{\gamma_t}}\right).
\end{align}
Due to the turbulence of phase-shift matrices, a closed-form OP expression for OMUR is intractable. Instead, we provide a relational result for performance comparison as follows:

\newtheorem{theorem}{Theorem}
\begin{theorem}
\label{Theorem_01}
The performance of OMUR is better than that of OR since its OP $P_{omur}(R_0)$ is less than $P_{or}(R_0)$.
\end{theorem}
\begin{IEEEproof}
Combining \eqref{eqn_E2EchannelGain} with \eqref{eqn_OMUR_gamma} gets $\gamma_{omur}=\gamma_{or}+\gamma_{e}$. The second term of \eqref{eqn_OMUR_gamma} is written as $\gamma_{e}=\sum_{k\neq k^\star} \left|\mathbf{f}^T \boldsymbol{\Theta}_{omur} \mathbf{g}_k +d_{k}\right|^2$ for notational simplicity. Since $\gamma_{e}>0$, we conclude
\begin{align} \nonumber \label{eqn:x}
    P_{omur}(R_0)&=\mathbb{P} \left\{ \log_2(1+ \gamma_{omur}\bar{\gamma_t})  < R_0 \right\}= \mathbb{P}(\gamma_{omur}<\gamma_0/\bar{\gamma_t}) \\ \nonumber
    &= \mathbb{P}(\gamma_{or}+\gamma_{e}<\gamma_0/\bar{\gamma_t})< \mathbb{P}(\gamma_{or}<\gamma_0/\bar{\gamma_t})=P_{or}(R_0).
\end{align}
\end{IEEEproof}

Last but not least, we are interested in analyzing ideal reflection (regardless of its non-existence in reality) for indicating the performance limit. It is known that the square of a Gamma RV follows the \textit{generalized Gamma} distribution \cite{Ref_sagias2005distribution}. Hence, the PDF of $\gamma_k=A_k^2$ is approximated as
\begin{equation} \label{ENQ_RIS_PDF_GGamma}
	f_{\gamma_k} (x) \approx 
	\frac{1}{2 \sqrt{x}} \frac{{\beta_k}^{\alpha_k}}{\Gamma(\alpha_k)} \! \left( \sqrt{x} \right)^{\alpha_k - 1} \! \! e^{-\beta_k \sqrt{x} }.
\end{equation}
Substitute \eqref{EQN_RIS_GammaK} into \eqref{IRS_EQN_SNR_ideal} to obtain $\gamma_{ir}=\sum_{k=1}^K \gamma_k$, namely the E2E channel power gain for IR is the \textit{sum} of generalized Gamma RVs. The exact distribution of $\gamma_{ir}$ is intractable. 
According to \cite{Ref_sagias2005distribution}, an upper bound for its CDF can be obtained through exploring another RV $Y=\prod_{k=1}^K \gamma_k$, namely the \textit{product} of generalized Gamma RVs. We have 
\begin{lemma}
The CDF of $Y$ is given by
\begin{align} \label{EQN_RIS_CDFofproductofGG}
    F_Y(x) &= \frac{V}{\sqrt{\lambda}}\left (\sqrt{2\pi} \right)^{\lambda-1}\\ \nonumber
    &\times G^{p,1}_{1,p+1} \left[  \frac{x^\lambda}{W\lambda^\lambda}\middle\vert \begin{array}{c}
1\\
\Delta(2\lambda;\alpha_1),\Delta(2\lambda;\alpha_2),\ldots,\Delta(2\lambda;\alpha_K),0
\end{array}   \right],
\end{align}
where $\Delta(a;b) \triangleq \frac{b}{a},\frac{b+1}{a},\ldots,\frac{b+(a-1)}{a} $, $G[\cdot]$ is the Meijer's G-function, $
V=\sqrt{\lambda}\Bigl (2\pi\Bigr)^{\frac{K+1-\lambda-p}{2}}\prod_{k=1}^K \frac{(2\lambda)^{\alpha_k-0.5}}{\Gamma(\alpha_k)}$,
and $    W=\frac{1}{\lambda^\lambda}\prod_{k=1}^K \left (\frac{2\lambda}{\beta_k}\right)^{2\lambda}$,
where $\lambda$ and $p$ are two positive integers satisfying $p = 2\lambda K$.
\end{lemma}
\begin{IEEEproof}
Compare \eqref{ENQ_RIS_PDF_GGamma} to obtain the parameters of (1) in \cite{Ref_sagias2005distribution} as $\beta_l=0.5$, $m_l=\alpha_k$ and $\Omega_l=\alpha_k/\beta_k$, $\forall l$. Substituting these parameters into \textit{Lemma 1} of \cite{Ref_sagias2005distribution} proves \eqref{EQN_RIS_CDFofproductofGG}. 
\end{IEEEproof}
Outage probability of IR is derived as 
\begin{align} \nonumber \label{eqn:xxxx}
    P_{ir}(R_0)&= \mathbb{P} \left\{ \log_2(1+ \gamma_{ir}\bar{\gamma_t} ) < R_0 \right\}\\ &=\mathbb{P}(\gamma_{ir}<\gamma_0/\bar{\gamma_t})
    =F_{\gamma_{ir}} (\gamma_0/\bar{\gamma_t}).
\end{align}
According to Theorem 1 of \cite{Ref_sagias2005distribution}, the CDF of $\gamma_{ir}$ is upper-bounded as $F_{\gamma_{ir}} (x)\leqslant F_Y((x/K)^K)$. 
Applying Lemma 2 and replacing $x=\left(\frac{\gamma_0}{K\bar{\gamma_t}}\right)^K$ into \eqref{EQN_RIS_CDFofproductofGG} yield
\begin{theorem}
\label{Theorem_02}
Outage probability of the ideal reflection has the following upper bound:
\begin{equation} \label{EQN_RIS_upperboundIDEAL}
    P_{ir}(R_0) \leqslant F_Y\left(\left(\frac{\gamma_0}{K\bar{\gamma_t}}\right)^K\right)
\end{equation}
\end{theorem}

\section{Numerical results} 

Monte-Carlo simulations are conducted to comparatively evaluate the performance of different schemes in a multi-RIS-aided system. 
Without loss of generality, we established the following simulation setup: the BS is located at the center of a circular cell with the radius of \SI{300}{\meter}, where $K=4$ users are  randomly distributed. Four surfaces with $N_s=100$ elements each are deployed equally on a concentric circle with a distance of \SI{60}{\meter} to the BS. The minimal UE power is set to $P_u=0.1\mathrm{W}$ with an incremental gain ranging from \SIrange{0}{30}{\decibel}. Signal bandwidth is $10\mathrm{MHz}$ and the noise power density equals $-174\mathrm{dBm/Hz}$ with noise figure  $9\mathrm{dB}$.
The large-scale fading for non-line-of-sight condition is computed according to the 3GPP Urban Micro (UMi) model as $\Omega = - 22.7 - 26 \log(f_c) - 36.7 \log(d)$, where $d$ is the distance and carrier frequency $f_c=2\mathrm{GHz}$. Considering the LOS component, the path loss of the BS-RIS link is calculated by $\mathcal{L}_0/d^{-\alpha}$,
where $\mathcal{L}_0=\SI{-30}{\decibel}$ is the path loss at the reference distance of \SI{1}{\meter} and the path-loss exponent $\alpha=2$. In addition, $m=2.5$ is applied for generating Nakagami-$m$ fading. 

\begin{figure}[!t]
\centering 
\subfloat[\scriptsize Four users $K=4$ for scheduling over independent Nakagami-m channels, where analytical results match numerical results.]{
\includegraphics[width=0.46\textwidth]{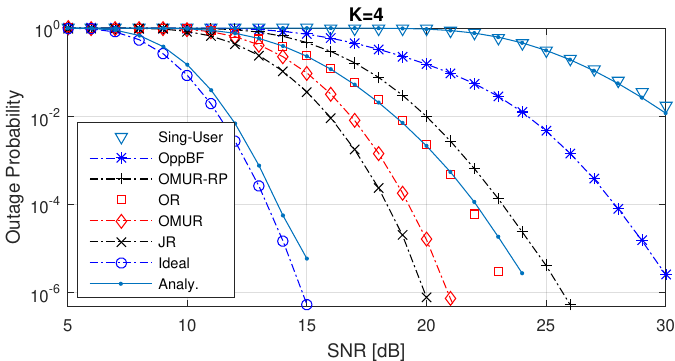}
\label{Fig_outagePerformance}
}
\\
\subfloat[\scriptsize Twenty users $K=20$ over either independent (solid lines) or correlated (dashed-dot lines) channels. Note that the curves of OR and OppBF are completely overlapped, the same as OMUR and OMUR-RP. ]{
\includegraphics[width=0.46\textwidth]{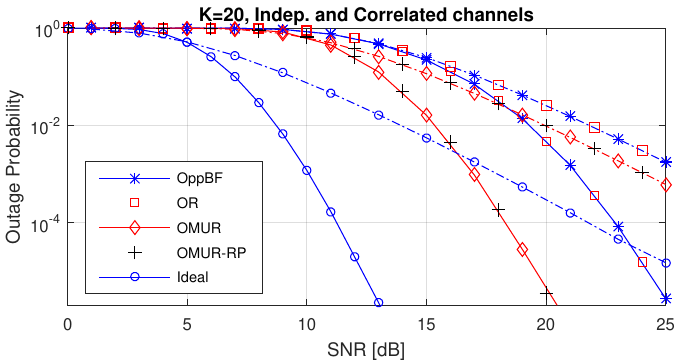}
\label{Fig_outageL20}
}
\caption{Performance comparison of different transmission schemes and settings in the uplink of a multi-RIS-aided multi-user system.}  
\end{figure}

\figurename \ref{Fig_outagePerformance} provides a comprehensive comparison among different schemes, including 1) the single-user case with its analytical result approximated by \eqref{Eqn:SU:OutageProb}; 2) OppBF proposed in \cite{Ref_nadeem2021opportunistic}; 3) OR and its analytical result approximated by \eqref{Eqn:OP_OpportuntisticReflection}; 4) OMUR; 5) OMUR-RP; 6) JR; and 7) IR with its upper bound given by \eqref{EQN_RIS_upperboundIDEAL}. It shows that the theoretical analysis is correct and the proposed scheme performs as expected. Since RISs can serve only one user optimally with its desired phase shifts, the gain of passive beamforming gradually vanishes as the increasing number of non-aligned users dominate the system. With a sufficiently large  number of  users, using random phase shifts is a more efficient choice. As shown in \figurename \ref{Fig_outageL20}, OR and OppBF achieve identical performance, as well as OMUR and OMUR-RP, under $K=20$ users. This figure also gives the numerical results over correlated channels using an inter-element correlation coefficient of 1, where the proposed scheme still outperforms OR and OppBF. 

Furthermore, the complexity is evaluated in terms of average CPU run time per channel realization. The platform uses an Intel i7-4790 CPU at 3.60GHz and 32GB memory, and runs Matlab parallel computing with 4 workers. As shown in Table I, due to the use of computation-hungry SDR approach, the complexity of JR is four orders of magnitude higher than that of the opportunistic methods. Considering the channel coherence time generally falls into the range of \SIrange{10}{100}{\milli \second}, JR is prohibitive for practical uses while OMUR is quite efficient.  
\begin{table}[!h]
\renewcommand{\arraystretch}{1.3}
\caption{Comparison of computational complexity.}
\label{table_complexity}
\centering
\begin{tabular}{|c|c|c|c|}
\hline
Schemes & \textit{Joint Reflection} & \textit{Opportunistic Reflection}&\textit{OMUR} \\
\hline \hline
CPU time [\si{\milli \second}] & $2.39\times 10^4$&  0.83&    1.42   \\ \hline
\end{tabular}
\end{table}

\section{Conclusions}
This paper proposed a novel multiple-access scheme named OMUR for multi-RIS-aided systems. By optimizing the RIS phase shifts for an opportunistic user, rather than all users, the reflection optimization is drastically simplified compared to joint reflection. Thanks to non-orthogonal transmission, the proposed scheme outperforms all single-user transmission schemes. A simplified version of OMUR exploits random phase shifts to avoid the high overhead of CSI acquisition in fast fading. It achieves the same performance as OMUR when the number of users is large. 
\bibliographystyle{IEEEtran} 
\bibliography{IEEEabrv,Ref_CommL}

\begin{thebibliography}{10}
\providecommand{\url}[1]{#1}
\csname url@samestyle\endcsname
\providecommand{\newblock}{\relax}
\providecommand{\bibinfo}[2]{#2}
\providecommand{\BIBentrySTDinterwordspacing}{\spaceskip=0pt\relax}
\providecommand{\BIBentryALTinterwordstretchfactor}{4}
\providecommand{\BIBentryALTinterwordspacing}{\spaceskip=\fontdimen2\font plus
\BIBentryALTinterwordstretchfactor\fontdimen3\font minus
  \fontdimen4\font\relax}
\providecommand{\BIBforeignlanguage}[2]{{%
\expandafter\ifx\csname l@#1\endcsname\relax
\typeout{** WARNING: IEEEtran.bst: No hyphenation pattern has been}%
\typeout{** loaded for the language `#1'. Using the pattern for}%
\typeout{** the default language instead.}%
\else
\language=\csname l@#1\endcsname
\fi
#2}}
\providecommand{\BIBdecl}{\relax}
\BIBdecl

\bibitem{Ref_jiang2023performance}
W.~Jiang and H.~Schotten, ``Performance impact of channel aging and phase noise
  on intelligent reflecting surface,'' \emph{{IEEE} Commun. Lett.}, vol.~27,
  no.~1, pp. 347--351, Jan. 2023.

\bibitem{Ref_jiang2023capacity}
W.~Jiang and H.~D. Schotten, ``Capacity analysis and rate maximization design
  in {RIS}-aided uplink multi-user {MIMO},'' in \emph{Proc. 2023 IEEE Wireless
  Commun. and Netw. Conf. (WCNC)}, Glasgow, Scotland, UK, Mar. 2023.

\bibitem{Ref_zheng2020intelligent_COML}
B.~Zheng, Q.~Wu, and R.~Zhang, ``Intelligent reflecting surface-assisted
  multiple access with user pairing: {NOMA or OMA}?'' \emph{{IEEE} Commun.
  Lett.}, vol.~24, no.~4, pp. 753 -- 757, Apr. 2020.

\bibitem{Ref_jiang2023orthogonal}
W.~Jiang and H.~Schotten, ``Orthogonal and non-orthogonal multiple access for
  intelligent reflection surface in {6G} systems,'' in \emph{Proc. 2023 IEEE
  Wireless Commun. and Netw. Conf. (WCNC)}, Glasgow, Scotland, UK, Mar. 2023.

\bibitem{Ref_ding2020simple}
Z.~Ding and H.~V. Poor, ``A simple design of {IRS-NOMA} transmission,''
  \emph{{IEEE} Commun. Lett.}, vol.~24, no.~5, pp. 1119 -- 1123, May 2020.

\bibitem{Ref_nadeem2021opportunistic}
Q.-U.-A. Nadeem \emph{et~al.}, ``Opportunistic beamforming using an intelligent
  reflecting surface without instantaneous {CSI},'' \emph{IEEE Wireless Commun.
  Lett.}, vol.~10, no.~1, pp. 146 -- 150, Jan. 2021.

\bibitem{Ref_jiang2023userscheduling}
W.~Jiang and H.~Schotten, ``User scheduling and passive beamforming for
  {FDMA/OFDMA} in intelligent reflection surface,'' in \emph{Proc. 2023 IEEE
  97th Veh. Techno. Conf. (VTC2023-Spring)}, Florence, Italy, Jun. 2023.

\bibitem{Ref_jiang2022multiuser}
------, ``Multi-user reconfigurable intelligent surface-aided communications
  under discrete phase shifts,'' in \emph{Proc. 36th IEEE Int. Workshop on
  Commun. Qual. and Reliability (CQR 2022)}, Arlington, United States, Sep.
  2022.

\bibitem{Ref_zheng2021doubleIRS}
B.~Zheng, C.~You, and R.~Zhang, ``Double-{IRS} assisted multi-user {MIMO}:
  Cooperative passive beamforming design,'' \emph{{IEEE} Trans. Wireless
  Commun.}, vol.~20, no.~7, pp. 4513 -- 4526, Jul. 2021.

\bibitem{Ref_niu2022double}
H.~Niu \emph{et~al.}, ``Double intelligent reflecting surface-assisted
  multi-user {MIMO} mmwave systems with hybrid precoding,'' \emph{{IEEE} Trans.
  Veh. Technol.}, vol.~71, no.~2, pp. 1575 -- 1587, Feb. 2022.

\bibitem{Ref_jiang2022intelligent}
W.~Jiang and H.~Schotten, ``Intelligent reflecting vehicle surface: A novel
  {IRS} paradigm for moving vehicular networks,'' in \emph{Proc. 2022 IEEE 40th
  Military Commun. Conf. (MILCOM 2022)}, Rockville, MA, USA, Nov. 2022.

\bibitem{Ref_jiang2023userselection}
------, ``User selection for simple passive beamforming in multi-ris-aided
  multi-user communications,'' in \emph{Proc. 2023 IEEE 97th Veh. Techno. Conf.
  (VTC2023-Spring)}, Florence, Italy, Jun. 2023.

\bibitem{Ref_wu2019intelligent}
Q.~Wu and R.~Zhang, ``Intelligent reflecting surface enhanced wireless network
  via joint active and passive beamforming,'' \emph{{IEEE} Trans. Wireless
  Commun.}, vol.~18, no.~11, pp. 5394 -- 5409, Nov. 2019.

\bibitem{cvx}
M.~Grant and S.~Boyd, ``{CVX}: Matlab software for disciplined convex
  programming, version 2.1,'' \url{http://cvxr.com/cvx}, Mar. 2014.

\bibitem{Ref_jiang20226G}
W.~Jiang and F.-L. Luo, \emph{6G Key Technologies: A Comprehensive
  Guide}.\hskip 1em plus 0.5em minus 0.4em\relax New York, USA: IEEE Press and
  John Wiley\&Sons, 2023, ch.~10.

\bibitem{Ref_do2021multiRIS}
T.~N. Do \emph{et~al.}, ``Multi-{RIS}-aided wireless systems: Statistical
  characterization and performance analysis,'' \emph{{IEEE} Trans. Commun.},
  vol.~69, no.~12, pp. 8641 -- 8658, Dec. 2021.

\bibitem{Ref_sagias2005distribution}
N.~Sagias \emph{et~al.}, ``On the distribution of the sum of generalized
  {Gamma} variates and applications to satellite digital communications,'' in
  \emph{Proc. 2nd Int. Symp. Wireless Commun. Sys.}, Siena, Italy, Sep. 2005.

\end{thebibliography}

\end{document}